\documentclass[sigconf]{acmart}

\usepackage{amsmath,amsfonts}
\usepackage{algorithmic}
\usepackage{graphicx}
\usepackage{textcomp}
\usepackage{xcolor}
\usepackage{tcolorbox}
\usepackage{graphicx}
\usepackage{listings}
\usepackage{makecell}
\usepackage{adjustbox}
\usepackage{multirow}
\usepackage{graphicx}
\usepackage{svg}
\usepackage{xspace}
\usepackage{amsmath}
\usepackage{subfigure}
\usepackage{caption}
\usepackage{amsmath}
\usepackage{booktabs}
\usepackage{color, colortbl}
\usepackage{url}
\usepackage{caption} 
\usepackage{pifont}
\usepackage{enumitem}
\usepackage{tabularx}

\newcolumntype{P}[1]{>{\centering\arraybackslash}p{#1}}

\DeclareUnicodeCharacter{2212}{-}
\definecolor{Gray}{gray}{0.9}
\definecolor{green}{RGB}{102,252,102}
\definecolor{ored}{RGB}{255,99,71}
\definecolor{orange}{RGB}{255,165,0}
\definecolor{lightgray}{RGB}{211,211,211}

\usepackage{xcolor}
\usepackage{xcolor,colortbl}
\definecolor{lightgray}{gray}{0.93}
\definecolor{slightgray}{gray}{0.98}
\definecolor{darkgray}{gray}{0.77}

\usepackage[ruled,vlined,linesnumbered]{algorithm2e}

\SetCommentSty{mycommfont}

\usepackage{tikz}
\usetikzlibrary{shapes}

\newcommand{\tool}{{\textsc{CAT}}\xspace}

\definecolor{amber}{rgb}{1.0, 0.49, 0.0}

\AtBeginDocument{%
  \providecommand\BibTeX{{%
    \normalfont B\kern-0.5em{\scshape i\kern-0.25em b}\kern-0.8em\TeX}}}

\setcopyright{acmcopyright}
\copyrightyear{2024}
\acmYear{2024}
\setcopyright{rightsretained}
\acmConference[ASE '24]{39th IEEE/ACM International Conference on Automated Software Engineering }{October 27-November 1, 2024}{Sacramento, CA, USA}
\acmBooktitle{39th IEEE/ACM International Conference on Automated Software Engineering (ASE '24), October 27-November 1, 2024, Sacramento, CA, USA}
\acmDOI{10.1145/3691620.3695260}
\acmISBN{979-8-4007-1248-7/24/10}

\begin{document}

\title{Enabling Cost-Effective UI Automation Testing with Retrieval-Based LLMs: A Case Study in WeChat}

\author{Sidong Feng}
\affiliation{%
  \institution{Monash University}
  \city{Melbourne}
  \country{Australia}
  }
\email{sidong.feng@monash.edu}

\author{Haochuan Lu}
\affiliation{%
  \institution{Tencent Inc.}
  \city{Guangzhou}
  \country{China}
  }
\email{hudsonhclu@tencent.com}

\author{Jianqin Jiang}
\affiliation{%
  \institution{Tencent Inc.}
  \city{Guangzhou}
  \country{China}
  }
\email{janetjiang@tencent.com}

\author{Ting Xiong}
\affiliation{%
  \institution{Tencent Inc.}
  \city{Guangzhou}
  \country{China}
  }
\email{candyxiong@tencent.com}

\author{Likun Huang}
\affiliation{%
  \institution{Tencent Inc.}
  \city{Guangzhou}
  \country{China}
  }
\email{likunhuang@tencent.com}

\author{Yinglin Liang}
\affiliation{%
  \institution{Tencent Inc.}
  \city{Guangzhou}
  \country{China}
  }
\email{dickylliang@tencent.com}

\author{Xiaoqin Li}
\affiliation{%
  \institution{Tencent Inc.}
  \city{Guangzhou}
  \country{China}
  }
\email{allysali@tencent.com}

\author{Yuetang Deng}
\affiliation{%
  \institution{Tencent Inc.}
  \city{Guangzhou}
  \country{China}
  }
\email{yuetangdeng@tencent.com}

\author{Aldeida Aleti}
\affiliation{%
  \institution{Monash University}
  \city{Melbourne}
  \country{Australia}
  }
\email{aldeida.aleti@monash.edu }

\renewcommand{\shortauthors}{Feng et al.}

\begin{abstract}
UI automation tests play a crucial role in ensuring the quality of mobile applications. Despite the growing popularity of machine learning techniques to generate these tests, they still face several challenges, such as the mismatch of UI elements. The recent advances in Large Language Models (LLMs) have addressed these issues by leveraging their semantic understanding capabilities. However, a significant gap remains in applying these models to industrial-level app testing, particularly in terms of cost optimization and knowledge limitation. To address this, we introduce \tool to create cost-effective UI automation tests for industry apps by combining machine learning and LLMs with best practices. Given the task description, \tool employs Retrieval Augmented Generation (RAG) to source examples of industrial app usage as the few-shot learning context, assisting LLMs in generating the specific sequence of actions. \tool then employs machine learning techniques, with LLMs serving as a complementary optimizer, to map the target element on the UI screen. Our evaluations on the WeChat testing dataset demonstrate the \tool's performance and cost-effectiveness, achieving 90\% UI automation with \$0.34 cost, outperforming the state-of-the-art. We have also integrated our approach into the real-world WeChat testing platform, demonstrating its usefulness in detecting 141 bugs and enhancing the developers' testing process.
\end{abstract}

\ccsdesc[500]{Software and its engineering~Software testing and debugging}

\keywords{UI automation test, large language model, retrieval-augmented generation, cost optimization}

\maketitle

\section{Introduction}

Mobile apps have become increasingly popular over the past decade, with millions of apps available for download from app stores like the Apple App Store and Google Play Store.
With the rise of app importance in our daily life, it has become increasingly critical for app developers to ensure that their apps are of high quality and perform as expected for users. 
One common practice for quality assurance is writing UI automation tests, enabling developers to use a pre-defined criterion as the test’s oracle to discover bugs or maximize code coverage.

Hardcoding UI automation tests, such as record and replay~\cite{zadgaonkar2013robotium,feng2023towards}, can quickly become obsolete due to the fast-paced evolution of platforms and the industry's demand for frequent releases. 
As a result, industry developers often favor writing UI automation tests with high-level task objectives, such as ``sharing a picture with Pony'' for regression or performance testing~\cite{choi2018detreduce}. 
The research community has made substantial contributions toward automating these testing activities. 
For instance, AppFlow~\cite{hu2018appflow} introduces a system that leverages test libraries of analogous apps and uses machine learning methods to synthesize UI automation tests. 
However, existing machine learning techniques in this field often face practical adoption challenges~\cite{feng2024mud}, particularly the issues of mismatched UI elements, which undermine the effectiveness of UI automation tests.
Large Language Models (LLMs) pave the way for numerous software development tasks, including UI automation. A recent success in this field is AdbGPT~\cite{feng2023prompting}, which uses prompt engineering to extract semantic understanding and logical reasoning from LLMs, significantly improving the performance of UI automation tests.


Yet, applying LLMs for UI automation tests in the industry involves two main challenges, in terms of cost optimization and knowledge integration. 
First, while numerous studies~\cite{feng2023prompting,feng2024mud} focus on utilizing LLMs for UI automation tests, the associated costs can be prohibitive for industry-level testing.
Even with the deployment of a private LLM service, the computational cost remains high. 
For instance, inferring a single token with LLaMA-7B~\cite{touvron2023llama} requires 6.7 billion FLOPs, and the entire UI automation process may use over millions of tokens. 
Second, LLMs often lack specific knowledge and experience regarding industrial apps, which can lead to ineffective UI automation.
This is primarily due to the potential license copyright violations from the LLMs~\cite{web:copyright}, which prevent the inclusion of industry resources in the training.

In this paper, we propose \tool, designed to facilitate \textbf{C}ost-effective UI \textbf{A}utomation \textbf{T}ests for industrial apps. 
This is achieved by integrating well-established machine learning methods with cutting-edge advancements in LLMs.
\tool operates in two phases: i) \textit{Task Description Decomposition} and ii) \textit{UI Automation Execution}. 
When presented with a task description, the first phase is to break down it into a specific sequence of executable actions.
Given the limited knowledge of industrial apps, we aim to provide one example to aid LLMs in comprehending the app usage and abstracting task description decomposition. 
As a result, we leverage the Retrieval Augmented Generation (RAG) method, which performs neural searches across previous app testing datasets to select the most analogous examples as the few-shot learning context, encouraging the LLMs to formulate possible actions.
Once the actions are determined, the second phase is to automatically execute them on the device by associating them with the UI element mapping. 
To accomplish this, we propose a machine learning method, with the LLMs serving as a complementary optimizer, to map the UI element on the screen.

To evaluate the performance of the \tool, we carry out a large-scale experiment involving 39k tasks in the WeChat dataset.
First, we conduct an ablation study to assess the performance of the \tool in UI automation, using five variations of the approach. 
The results show that \tool significantly increases the completion rate at a reduced cost, successfully completing 90\% of the tasks at an average cost of \$0.34. 
Second, we evaluate the performance of the \tool against two state-of-the-art UI automation methods. 
Our findings reveal that our approach can save more cost and time without sacrificing the completion rate, saving \$1,467 cost even when compared to the best baseline. 
Beyond the performance of the approach, we also evaluate its practical usefulness. 
We integrate the \tool with the WeChat testing platform, triggering it whenever new functionalities need testing or a new version is released, accompanied by new task descriptions. 
During the testing period from December 2023 to June 2024, \tool automatically executes 6k of UI automation tests, detecting 141 bugs. 
This alleviates the developers' burden in bug detection, saving substantial time for subsequent bug fixing.

The contributions of this paper are as follows:
\begin{itemize}
    \item We present \tool, a novel approach that combines machine learning and large language models with the adoption of best practices to generate UI automation tests. To the best of our knowledge, this is the first work that contemplates cost optimization and knowledge integration of LLMs in industrial-level app testing.
    \item We conduct extensive experiments, including ablation studies and comparisons with state-of-the-art methods, to showcase the performance and cost-effectiveness of \tool. Additionally, we incorporate our approach into a real-world testing platform to illustrate its practical usefulness.
\end{itemize}

\section{Approach}
\begin{figure}
	\centering
	\includegraphics[width = 0.8\linewidth]{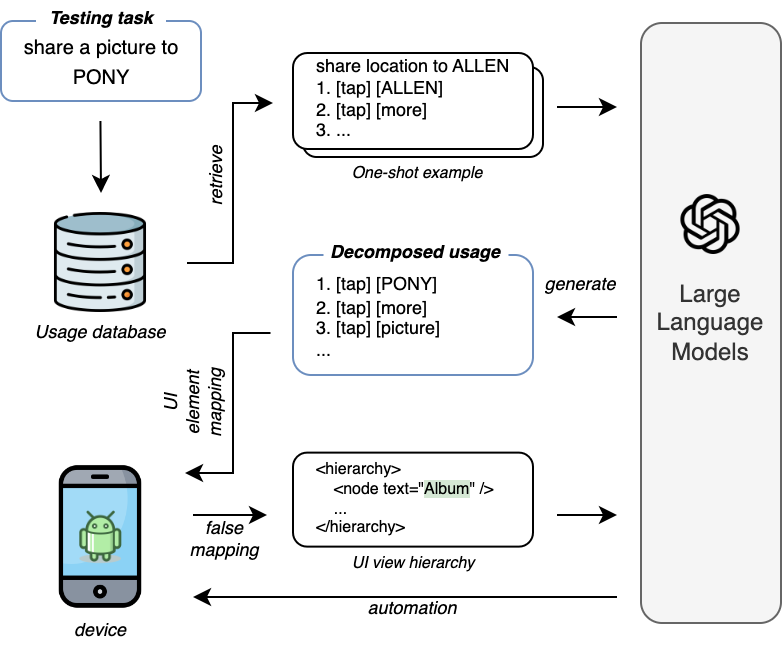}
	\caption{The overview of our approach.}
	\label{fig:overview}
\end{figure}

Given a high-level task description, we propose an automated approach to break down the actions and execute them on the dynamic UI to trigger the app activity for testing.
The overview of our approach \tool is shown in Figure~\ref{fig:overview}, which is divided into two main phases: (i) the \textit{Task Description Decomposition} phase, which decomposes task description into multiple potential action steps, including action types, target elements, and input values;
(ii) the \textit{UI Automation Execution} phase which aligns the actions with the dynamic UI elements to accomplish UI automation.

\subsection{Task Description Decomposition}
\label{sec:phase1}
The first phase of our approach is to understand, analyze, organize, and generate the actions from the high-level task description by using the advance of LLMs.
Given the potential knowledge gap in LLMs regarding the specifics of industrial app usage, we adopt the widely recognized Retrieval Augmented Generation (RAG) technique. 
It first employs a machine learning mechanism to retrieve a few examples detailing the inputs and outputs associated with industrial app usage.
Then, these examples serve as a context for few-shot learning, assisting the LLMs in generating potential actions for the task description.

\subsubsection{Retrieving Few-shot Examples}
\label{sec:retrival-based}
A representative example helps the model in eliciting specific knowledge and abstractions.
A randomly chosen example might not capture the intricacies of industrial app usage, thereby limiting the LLMs' ability to fully comprehend the task. 
To address this, we utilize examples from previous app testing datasets, a common industry practice for testing app functionalities across iterative versions. 
The dataset usually comprises testing objectives with high-level task descriptions and their specific sequence of steps.
Consequently, we apply a neural search and a similarity-based retrieval technique to select representative examples from the dataset.


First, we utilize the transformer encoder model T5~\cite{raffel2020exploring} to encode the descriptions into a vector space.
This allows for the handling of variable-sized inputs: when an input sequence is provided, it is mapped to a sequence of embeddings that are then passed into the encoder. 
All encoders share an identical structure, each consisting of two subcomponents: a self-attention layer followed by a compact feed-forward network. 
Layer normalization is applied to each subcomponent's input, while a residual skip connection adds each subcomponent's input to its output. 
Dropout is implemented within the feed-forward network, on the skip connection, on the attention weights, and at the input and output of the entire stack.

Based on the vectors, we employ the sparse retrieval method known as cosine similarity~\cite{singhal2001modern}, to identify representative examples that align closely with the testing task description, as indicated by the highest relevance score.
In detail, it is measured as: 
\(cos(v_1, v_2) = \frac{v_{1} v_{2}}{\left | v_1 \right |\left | v_2 \right |} \), where $v_{1} v_{2}$ is the inner product of two vectors and $\left | v_1 \right |\left | v_2 \right |$ is the product of 2-norm for these vectors. 


\begin{table}
        \footnotesize
	\centering
	\caption{Prompt example for task description decomposition. }
	\label{tab:prompt1}
	\begin{tabularx}{\linewidth}{l|X} 
		\hline
		\rowcolor{darkgray} \bf{COMPONENT}  & \bf{DETAILS} \\
		\hline
		 ⟨Instructions⟩ & \textit{\makecell[Xt]{I want you to act as a professional developer. I expect you to analyze the task description I provide and respond with potential actions that could interact with the device. Please enumerate these actions and encapsulate the operations within brackets [like this].}} \\
		  \hline
	⟨One-shot Example⟩ & \textit{\makecell[Xt]{Here is an example: \\ \hspace{0.15cm} Task description -> ``open settings'' \\ \hspace{0.15cm} Actions -> 1. [tap] [me] 2. [tap] [settings] ...}} \\
		  \hline
		   ⟨Testing Task Description⟩ & \textit{\makecell[Xt]{Here is the testing task: \\ \hspace{0.15cm} Task description -> ``change username to TEST'' \\ \hspace{0.15cm} Actions -> }} \\
            \hline
	\end{tabularx}
\end{table}

\subsubsection{Generating Decomposed Actions}
A LLMs prompt example to generate potential actions for task description is shown in Table~\ref{tab:prompt1}, including \texttt{⟨Instructions⟩} + \texttt{⟨One-shot Example⟩} + \texttt{⟨Testing Task Description⟩}.
Specifically, we first instruct the LLMs to outline the objective goal, which is to dissect the task description into the potential sequence of actions.
According to a small pilot study, we retrieve the top-1 representative examples for few-shot learning (as known as one-shot learning), aiding in the recognition of industrial app usage and output patterns.
Next, we present the testing task description as the test prompt and ask for the decomposed actions.
Due to the advantage of instruction prompting and few-shot learning, the LLMs will consistently generate a numeric list to represent the sequence of actions in the same format as our example output, which can be inferred using regular expressions.


        

\subsection{UI Automation Execution}
\label{sec:phase2}
The second phase involves matching the decomposed actions with the UI elements to automate execution. 
A common approach is to use lexical computation to match the displayed text of the elements on the current UI screen. 
However, there may be mismatches with the target element due to the dynamic nature of the UI screen. 
For instance, ``sharing a moment'' might be synthesized into a ``moment A'' element due to previous app usage, while the current element is ``moment B'' on the UI screen. 
This can hinder the process of machine learning-based UI element mapping, leading to UI automation failure.
To address this issue, we utilize the ability of LLMs to semantically explore the app and understand the semantic correlation towards the target element. 
Note that, we use the machine learning method as the primary UI element mapping and employ the LLMs as a complementary optimizer to save expenditure costs for industrial-level testing.

\subsubsection{UI Element Mapping}
Consider the target element ($e_{target}$) in the executive actions and the elements in the current UI screen ($\{e_1, e_2, ..., e_n\}$).
Similar to the previous methods in Section~\ref{sec:retrival-based}, we first use the transformer-based model to encode the UI elements into vectors.
We then use the similarity measurement method to compute the lexical similarity between UI elements and the target element: \(Similarity(e_{n}, e_{target})) = cos(Encode(e_{n}), Encode(e_{target}))\).
To match the most similar UI element on the UI screen to the target element, we identify the UI element with the highest similarity value.
Additionally, we establish a threshold value to ascertain whether the target element is matched, or if it may contain semantic mismatches, which would necessitate optimization by LLMs.

\begin{table}
        \footnotesize
	\centering
	\caption{Prompt example for UI element mapping. }
	\label{tab:prompt2}
	\begin{tabularx}{\linewidth}{l|X} 
		\hline
		\rowcolor{darkgray} \bf{COMPONENT}  & \bf{DETAILS} \\
		\hline
		 ⟨Instructions⟩ & \textit{\makecell[Xt]{I want you to act as a professional developer. I would like you to analyze a given target element and the current UI screen, provided in the format of an XML view hierarchy. Please identify and respond with an element on the UI screen within brackets [like this], that is semantically related to the target element.}} \\
		  \hline
		   ⟨Testing Element⟩ & \textit{\makecell[Xt]{Target element -> ``username''\\ Current UI screen -> <view hierarchy> \\ Identified element -> }} \\
            \hline
	\end{tabularx}
\end{table}

\subsubsection{LLMs Optimizer}
LLMs are employed to specifically address occasional mismatches in UI elements.
Table~\ref{tab:prompt2} presents an example of the prompts given to LLMs. 
The prompt begins with an instruction outlining the objective of identifying UI elements on the screen that are semantically related to the target element.
A challenge in using LLMs for UI element mapping is their limitation in processing large text inputs, while the UI representation (i.e., view hierarchy) is typically lengthy - averaging thousands of tokens for each UI.
Although recent LLMs can handle visual inputs such as UI screens, recent research~\cite{bubeck2023sparks} has highlighted constraints in visual UI understanding.
To address this, we propose a heuristic method to simplify nested layouts and extract atomic elements by traversing the view hierarchy tree using a depth-first search. 
Specifically, we iterate through each node, starting from the root of the view hierarchy, and remove layouts that contain only one node, continuing to search its child node. 
With the simplified UI representation, we prompt the LLMs to identify which UI element is semantically related to the target element. 

\subsection{Implementation}
\label{sec:phase3}
Our \tool is implemented as a fully automated UI automation testing tool.
According to a small-scale pilot study, we use the pre-trained ChatGPT model as the LLMs which was released on the OpenAI website~\cite{web:chatgpt}.
The basic model of ChatGPT is the gpt-4 model, representing the state-of-the-art LLMs.
To execute the UI automation on the device, we use Genymotion~\cite{web:genymotion} for running and controlling the virtual Android device, Android UIAutomator~\cite{web:uiautomator} for dumping the UI view hierarchy, and Android Debug Bridge (ADB)~\cite{web:adb} for executing the operations.


\section{Evaluation}
\label{sec:evaluation}
In this section, we describe the procedure we used to evaluate \tool.

\begin{itemize}[leftmargin=0.3cm] 
    \item \textbf{RQ1:} How effective is our approach in UI automation?
    \item \textbf{RQ2:} How does our approach compare to state-of-the-art?
    \item \textbf{RQ3:} How useful is our approach in a real-world testing environment?
\end{itemize}

\subsection{RQ1: Performance of UI Automation}
\label{sec:rq1}
\textbf{Experimental Setup.}
We collect 39,981 task descriptions from the WeChat app\footnote{WeChat is among the most popular messenger apps in the world with over 1.67 billion monthly active users.} as our experimental dataset.
Each description, averaging 18.7 words, is provided by 71 internal developers and each invokes a core functionality of the app (on average 7.3 action steps).
These descriptions and their corresponding ground-truth action steps are utilized for UI automation tests across 24 app development cycles over a span of one year.
We divide the experimental dataset into testing and retrieval datasets for our evaluation.
Note that a simple random split may not adequately evaluate generalizability, as the same activities may have very similar descriptions.
To avoid this data leakage problem~\cite{kaufman2012leakage}, we partition the descriptions in the dataset by app activities.
The resulting split includes 2,010 (5\%) descriptions in the testing dataset, and 37,971 (95\%) in the retrieval dataset.

\textbf{Baselines.}
We set up four ablation studies as baselines to compare with our approach.
Given that the \tool comprises two main phases, we perform variations of our approach for each phase.
In Section~\ref{sec:phase1}, we introduce a retrieval-based method to select the top-1 examples to be used in the prompt for few-shot learning.
Thus, we examine the search space of no examples (\textit{0-shot RAG}) and N selected examples (\textit{N-shot RAG}).

In Section~\ref{sec:phase2}, we present an optimizer that utilizes LLMs to address the issues of mismatched UI elements on the dynamic UI screen.
Thus, we consider a variant of \textit{\tool (no optimizer)} to compare the performance of our approach with and without LLMs as the complementary method.

In Section~\ref{sec:phase3}, we detail the implementation of our approach, employing ChatGPT as the LLMs. As a variant, we set up an ablation study using the open-sourced LLMs LLaMA70B~\cite{touvron2023llama}, referred to as \textit{\tool (LLaMA70B)}.

\textbf{Evaluation Metrics.}
We employ three evaluation metrics: completion rate, financial cost, and time overhead. 
The completion rate assesses the ability of the approach to successfully automate the task within the app. 
A higher completion rate indicates a more effective approach to executing UI automation tests. 
As the ultimate goal is cost-effectiveness, we calculate the average expenditure incurred for inferring the LLMs throughout the UI automation tests. 
Lastly, we also measure the duration of time spent, in minutes.

\textbf{Results.}
Table~\ref{tab:rq1_performance} illustrates the performance of \tool in executing UI automation for task descriptions.
Our approach achieves an average completion rate of 90\% at a cost of \$0.34, outperforming the ablation baselines.
We observe that the method excluding few-shot learning (0-shot RAG) only attains a 50\% completion rate due to the reason that the LLMs' lack of knowledge and experience with certain apps.
In comparison to 0-shot RAG, the use of few-shot examples can significantly help LLMs in understanding app usage knowledge, increasing the completion rate by 40\% for 1-shot learning in our approach.
However, supplementing extensive examples (N-shot RAG) does not enhance the approach's performance. 
This is because the longer context of the examples might lead the LLMs to display tendencies towards instruction forgetting, format errors, and abnormal reasoning.
Instead, it incurs a much higher cost, making it 150\% more expensive than our approach.

In addition, optimizing UI element mapping using LLMs can further improve the approach's performance of 38\% completion rate compared to the ablation baseline of \tool without an optimizer.
This suggests that the LLMs possess the ability to comprehend the semantic correlation between the target elements and the elements in the dynamic UI, potentially identifying the likely operations.

Despite the fact that using open-sourced LLMs does not involve financial costs, the state-of-the-art model LLaMA70B only achieves a 71\% completion rate. 
However, we are optimistic that once these open-sourced LLMs can provide comparable performance with the ChatGPT model, we could integrate our base model with the open-sourced LLMs, thereby further reducing costs for future UI automation tests.

\begin{table}
    \small
	\centering
	\caption{Performance comparison of ablation studies.}
	\label{tab:rq1_performance}
	\begin{tabular}{l||c|c|c} 
	    \hline
	   \bf{Method} & \bf{Completion} & \bf{Avg. Cost} & \bf{Avg. Time}\\
	    \hline
            \tool (0-shot RAG) & 50\% & \$0.34 & 2.61 min \\
		\hline
            \tool (N-shot RAG) & 66\% & \$0.85 & 3.20 min \\
            \hline
        \tool (no optimizer) & 52\% & \$0.01 & 5.26 min \\
        \hline
        \tool (LLaMA70B) & 71\% & - & 4.07 min \\
        \hline
        \hline
        \rowcolor{darkgray} \tool & 90\% & \$0.34 & 2.65 min \\
            \hline
	\end{tabular}
\end{table}

\subsection{RQ2: Comparison with State-of-the-Art}
\label{sec:rq2}
\textbf{Experimental Setup.}
To answer RQ2, we evaluate the comparison of our approach to the state-of-the-art baselines. 
We also use the experimental dataset collected in RQ1 (Section~\ref{sec:rq1}).

\textbf{Baselines.}
We set up two state-of-the-art methods as baselines for comparison with our approach.
These include two machine-learning-based methods and one LLMs-based method, all of which are widely used for UI automation.
\textit{Seq2Act}~\cite{li2020mapping} utilizes models to extract actions and object targets from task descriptions, associating them with UI elements to facilitate UI automation.
\textit{AdbGPT}~\cite{feng2023prompting} employs the recent advancements of LLMs for automating UI tasks to extract action entities from task descriptions and make decisions for selecting executable UI elements.

\textbf{Evaluation Metrics.}
To compare with the state-of-the-art baselines, we also employ three evaluation metrics, i.e., completion rate, financial cost, and time overhead.

\textbf{Results.}
Table~\ref{tab:rq2_performance} presents the performance comparison with the baselines. 
The machine learning-based method (Seq2Act), while not incurring any financial costs, only attains completion rates of 35\%, which is 55\% lower than our approach \tool. 
Among the baselines, AdbGPT exhibits the best performance, achieving a completion rate of 90\%. However, it is costly, averaging \$1.07 for UI automation due to the extensive use of LLMs. 
In contrast, our approach, \tool, which integrates machine-learning methods and LLMs optimally, saves \$1,467 without compromising the completion rate.

\begin{table}
    \small
	\centering
	\caption{Performance comparison of state-of-the-art.}
	\label{tab:rq2_performance}
	\begin{tabular}{l||c|c|c} 
	    \hline
	    \bf{Method} & \bf{Completion} & \bf{Avg. Cost} & \bf{Avg. Time}\\
	     \hline
            Seq2Act~\cite{li2020mapping} & 35\% & - & 5.89 min \\
            \hline
            AdbGPT~\cite{feng2023prompting} & 90\% & \$1.07 & 2.91 min \\
            \hline
            \rowcolor{darkgray} \tool & 90\% & \$0.34 & 2.63 min \\
            \hline
	\end{tabular}
\end{table}


\subsection{RQ3: Usefulness of \tool}
\label{sec:rq3}
\textbf{Industrial Usage.}
We collaborate with Tencent to incorporate our approach, \tool, into the WeChat testing platform. 
\tool is integrated into the internal UI automation process and is activated whenever new features are proposed or a new app version is released, using the new task descriptions. 
As the ultimate goal aim of UI automation is to detect bugs, we also combine our approach with internal bug detection methods through UI automation tests.
Specifically, we use heuristics to monitor event logs to identify crash bugs. 
For non-crash bugs, we leverage previous studies such as UI display bug detection~\cite{xie2020uied,xie2022psychologically,chen2019gallery,feng2022gallery} and functional bug detection~\cite{chen2023unveiling,feng2021auto,feng2022autop}.
We set up the number of bugs as the evaluation metric to assess the usefulness of our approach.

\textbf{Results.}
We run the experiment in the WeChat testing system, with over 6,300 task descriptions from December 2023 to June 2024. 
The completion rate for UI automation reached 88\%, indicating that most task descriptions can be successfully automated and tested without the need for developer interaction. 
In addition, we automatically detect 141 bugs in these task descriptions. 
Note that the detected bugs are also reviewed by internal developers to verify their validity. 
This indicates the usefulness of our \tool in reducing human effort in bug detection and significantly saving time in subsequent bug fixes.
\section{Implication and Discussion}

Although many tools~\cite{feng2023prompting,wang2024feedback,feng2024mud} have been proposed that rely solely on LLMs for UI automation, we have opted to combine mature machine learning with LLMs for industrial-level app testing, based on two practical lessons learned.
First, the computational cost for LLMs is high, which is not feasible in an industry that often requires hundreds or thousands of testing tasks.
Second, the LLMs are not entirely reliable. 
Due to the nature of generative models, LLMs may often hallucinate, i.e., respond with unrelated concepts. In contrast, a hybrid of machine learning methods and LLMs can significantly mitigate these issues, making the testing system more robust for industrial use.




Another potential interest is the generalizability of our approach to other industrial apps.
In this study, we focus on the WeChat app, using the 39k tasks in the WeChat dataset to assess the performance of our approach.
We believe that our approach should be easily adapted to other industrial apps, given the commonality of the dataset used in industrial app testing~\cite{gao2019emerging}.
\section{Related Work}

Existing automated test generation techniques~\cite{gu2019practical,su2017guided,web:monkey,mao2016sapienz} share a complementary objective to ours: they mainly focus on generating tests to maximize code coverage and detect bugs.
As opposed, we aim to generate tests that cover specific functionalities (similar to our definition of tasks), guided by manually written descriptions.
The area of written descriptions that most closely aligns with test generation is the work on script-based record and replay.
For example, RERAN~\cite{gomez2013reran} is among the earliest record-and-replay tools utilizing the Linux kernel.
Guo et al.~\cite{guo2019sara} introduce an industrial-scale record-and-replay tool, SARA, designed for widget-sensitive and time-sensitive recording and replaying.
Over the past few years, several tools like Robotium~\cite{zadgaonkar2013robotium}, and WeReplay~\cite{feng2023towards,feng2023efficiency}, have been developed.
Additionally, there are numerous supplementary tools~\cite{feng2023read,feng2022gifdroid,feng2022gifdroid1,feng2023video2action} that assist in script writing.
However, these scripts often refer to UI elements with absolute positions or rely on fragile rules, which hinders industries from adopting these methods.

Consequently, many studies aim to advance upon this by simplifying the process, relying on high-level natural language descriptions that outline the desired test tasks.
For instance, AppFlow~\cite{hu2018appflow} synthesizes UI automation tests according to the test library.
However, these synthesized tests may lack robustness and contain specific issues like mismatched UI elements, leading to test failure.
Recently, some studies have utilized LLMs to address these failures.
For example, Feng et al.~\cite{feng2023prompting} introduce AdbGPT which equips the LLMs with the semantic understanding to guide the replay of certain descriptions with impressive performance.
However, using LLMs for these stochastic app explorations might be expensive and ineffective for industrial app testing due to cost optimization and knowledge integration.
In contrast, we propose a hybrid approach, \tool, that combines machine learning methods and LLMs to adopt the best practice for generating robust UI automation tests for cost-effective industrial-level testing.


\section{Conclusion}
We present \tool, a practical solution for generating UI automation tests for mobile apps. 
Specifically, given a high-level task description, \tool first applies RAG to retrieve relevant app usages as examples to elicit specific knowledge for LLMs to generate a concrete sequence of actions. 
It then employs machine learning and LLMs to adapt these actions to the app's dynamic UI, correcting any element discrepancies.
Our evaluation of the WeChat dataset, comprising 39k tasks, shows a 90\% completion rate in executing UI automation tests at just \$0.34 per test.
Additionally, its integration into the WeChat testing platform has led to the automatic identification of 141 bugs, easing the test-writing burden for developers.


\begin{acks}
We appreciate the assistance from Chunyang Chen and the WeCom team for their valuable contributions to the methodology discussion and some experimental processes. This research is supported by the Australian Research Council under grant DP210100041.
\end{acks}

\bibliographystyle{ACM-Reference-Format}
\bibliography{main.bib}


\begin{thebibliography}{37}


\ifx \showCODEN    \undefined \def \showCODEN     #1{\unskip}     \fi
\ifx \showDOI      \undefined \def \showDOI       #1{#1}\fi
\ifx \showISBNx    \undefined \def \showISBNx     #1{\unskip}     \fi
\ifx \showISBNxiii \undefined \def \showISBNxiii  #1{\unskip}     \fi
\ifx \showISSN     \undefined \def \showISSN      #1{\unskip}     \fi
\ifx \showLCCN     \undefined \def \showLCCN      #1{\unskip}     \fi
\ifx \shownote     \undefined \def \shownote      #1{#1}          \fi
\ifx \showarticletitle \undefined \def \showarticletitle #1{#1}   \fi
\ifx \showURL      \undefined \def \showURL       {\relax}        \fi
\providecommand\bibfield[2]{#2}
\providecommand\bibinfo[2]{#2}
\providecommand\natexlab[1]{#1}
\providecommand\showeprint[2][]{arXiv:#2}

\bibitem[web(2024a)]%
        {web:adb}
 \bibinfo{year}{2024}\natexlab{a}.
\newblock \bibinfo{title}{Android Debug Bridge (adb) - Android Developers}.
\newblock \bibinfo{howpublished}{\url{https://developer.android.com/studio/command-line/adb}}.
\newblock


\bibitem[web(2024b)]%
        {web:uiautomator}
 \bibinfo{year}{2024}\natexlab{b}.
\newblock \bibinfo{title}{Android Uiautomator2 Python Wrapper}.
\newblock \bibinfo{howpublished}{\url{https://github.com/openatx/uiautomator2}}.
\newblock


\bibitem[web(2024c)]%
        {web:copyright}
 \bibinfo{year}{2024}\natexlab{c}.
\newblock \bibinfo{title}{Developers warned: GitHub Copilot code may be licensed}.
\newblock \bibinfo{howpublished}{\url{https://www.techtarget.com/searchsoftwarequality/news/252526359/Developers-warned-GitHub-Copilot-code-may-be-licensed}}.
\newblock


\bibitem[web(2024d)]%
        {web:genymotion}
 \bibinfo{year}{2024}\natexlab{d}.
\newblock \bibinfo{title}{Genymotion – Android Emulator for app testing}.
\newblock \bibinfo{howpublished}{\url{https://www.genymotion.com/}}.
\newblock


\bibitem[web(2024e)]%
        {web:chatgpt}
 \bibinfo{year}{2024}\natexlab{e}.
\newblock \bibinfo{title}{Introducing ChatGPT}.
\newblock \bibinfo{howpublished}{\url{https://chat.openai.com/}}.
\newblock


\bibitem[web(2024f)]%
        {web:monkey}
 \bibinfo{year}{2024}\natexlab{f}.
\newblock \bibinfo{title}{UI/Application Exerciser Monkey}.
\newblock \bibinfo{howpublished}{\url{https://developer.android.com/studio/test/other-testing-tools/monkey}}.
\newblock


\bibitem[Bubeck et~al\mbox{.}(2023)]%
        {bubeck2023sparks}
\bibfield{author}{\bibinfo{person}{S{\'e}bastien Bubeck}, \bibinfo{person}{Varun Chandrasekaran}, \bibinfo{person}{Ronen Eldan}, \bibinfo{person}{Johannes Gehrke}, \bibinfo{person}{Eric Horvitz}, \bibinfo{person}{Ece Kamar}, \bibinfo{person}{Peter Lee}, \bibinfo{person}{Yin~Tat Lee}, \bibinfo{person}{Yuanzhi Li}, \bibinfo{person}{Scott Lundberg}, {et~al\mbox{.}}} \bibinfo{year}{2023}\natexlab{}.
\newblock \showarticletitle{Sparks of artificial general intelligence: Early experiments with gpt-4}.
\newblock \bibinfo{journal}{\emph{arXiv preprint arXiv:2303.12712}} (\bibinfo{year}{2023}).
\newblock


\bibitem[Chen et~al\mbox{.}(2019)]%
        {chen2019gallery}
\bibfield{author}{\bibinfo{person}{Chunyang Chen}, \bibinfo{person}{Sidong Feng}, \bibinfo{person}{Zhenchang Xing}, \bibinfo{person}{Linda Liu}, \bibinfo{person}{Shengdong Zhao}, {and} \bibinfo{person}{Jinshui Wang}.} \bibinfo{year}{2019}\natexlab{}.
\newblock \showarticletitle{Gallery dc: Design search and knowledge discovery through auto-created gui component gallery}.
\newblock \bibinfo{journal}{\emph{Proceedings of the ACM on Human-Computer Interaction}} \bibinfo{volume}{3}, \bibinfo{number}{CSCW} (\bibinfo{year}{2019}), \bibinfo{pages}{1--22}.
\newblock


\bibitem[Chen et~al\mbox{.}(2023)]%
        {chen2023unveiling}
\bibfield{author}{\bibinfo{person}{Jieshan Chen}, \bibinfo{person}{Jiamou Sun}, \bibinfo{person}{Sidong Feng}, \bibinfo{person}{Zhenchang Xing}, \bibinfo{person}{Qinghua Lu}, \bibinfo{person}{Xiwei Xu}, {and} \bibinfo{person}{Chunyang Chen}.} \bibinfo{year}{2023}\natexlab{}.
\newblock \showarticletitle{Unveiling the tricks: Automated detection of dark patterns in mobile applications}. In \bibinfo{booktitle}{\emph{Proceedings of the 36th Annual ACM Symposium on User Interface Software and Technology}}. \bibinfo{pages}{1--20}.
\newblock


\bibitem[Choi et~al\mbox{.}(2018)]%
        {choi2018detreduce}
\bibfield{author}{\bibinfo{person}{Wontae Choi}, \bibinfo{person}{Koushik Sen}, \bibinfo{person}{George Necula}, {and} \bibinfo{person}{Wenyu Wang}.} \bibinfo{year}{2018}\natexlab{}.
\newblock \showarticletitle{DetReduce: minimizing Android GUI test suites for regression testing}. In \bibinfo{booktitle}{\emph{Proceedings of the 40th International Conference on Software Engineering}}. \bibinfo{pages}{445--455}.
\newblock


\bibitem[Feng and Chen(2022a)]%
        {feng2022gifdroid1}
\bibfield{author}{\bibinfo{person}{Sidong Feng} {and} \bibinfo{person}{Chunyang Chen}.} \bibinfo{year}{2022}\natexlab{a}.
\newblock \showarticletitle{GIFdroid: an automated light-weight tool for replaying visual bug reports}. In \bibinfo{booktitle}{\emph{Proceedings of the ACM/IEEE 44th International Conference on Software Engineering: Companion Proceedings}}. \bibinfo{pages}{95--99}.
\newblock


\bibitem[Feng and Chen(2022b)]%
        {feng2022gifdroid}
\bibfield{author}{\bibinfo{person}{Sidong Feng} {and} \bibinfo{person}{Chunyang Chen}.} \bibinfo{year}{2022}\natexlab{b}.
\newblock \showarticletitle{Gifdroid: Automated replay of visual bug reports for android apps}. In \bibinfo{booktitle}{\emph{Proceedings of the 44th International Conference on Software Engineering}}. \bibinfo{pages}{1045--1057}.
\newblock


\bibitem[Feng and Chen(2024)]%
        {feng2023prompting}
\bibfield{author}{\bibinfo{person}{Sidong Feng} {and} \bibinfo{person}{Chunyang Chen}.} \bibinfo{year}{2024}\natexlab{}.
\newblock \showarticletitle{Prompting is all you need: Automated android bug replay with large language models}. In \bibinfo{booktitle}{\emph{Proceedings of the 46th IEEE/ACM International Conference on Software Engineering}}. \bibinfo{pages}{1--13}.
\newblock


\bibitem[Feng et~al\mbox{.}(2022a)]%
        {feng2022gallery}
\bibfield{author}{\bibinfo{person}{Sidong Feng}, \bibinfo{person}{Chunyang Chen}, {and} \bibinfo{person}{Zhenchang Xing}.} \bibinfo{year}{2022}\natexlab{a}.
\newblock \showarticletitle{Gallery dc: Auto-created gui component gallery for design search and knowledge discovery}. In \bibinfo{booktitle}{\emph{Proceedings of the ACM/IEEE 44th International Conference on Software Engineering: Companion Proceedings}}. \bibinfo{pages}{80--84}.
\newblock


\bibitem[Feng et~al\mbox{.}(2023a)]%
        {feng2023video2action}
\bibfield{author}{\bibinfo{person}{Sidong Feng}, \bibinfo{person}{Chunyang Chen}, {and} \bibinfo{person}{Zhenchang Xing}.} \bibinfo{year}{2023}\natexlab{a}.
\newblock \showarticletitle{Video2Action: Reducing human interactions in action annotation of app tutorial videos}. In \bibinfo{booktitle}{\emph{Proceedings of the 36th Annual ACM Symposium on User Interface Software and Technology}}. \bibinfo{pages}{1--15}.
\newblock


\bibitem[Feng et~al\mbox{.}(2022b)]%
        {feng2022autop}
\bibfield{author}{\bibinfo{person}{Sidong Feng}, \bibinfo{person}{Minmin Jiang}, \bibinfo{person}{Tingting Zhou}, \bibinfo{person}{Yankun Zhen}, {and} \bibinfo{person}{Chunyang Chen}.} \bibinfo{year}{2022}\natexlab{b}.
\newblock \showarticletitle{Auto-icon+: An automated end-to-end code generation tool for icon designs in ui development}.
\newblock \bibinfo{journal}{\emph{ACM Transactions on Interactive Intelligent Systems}} \bibinfo{volume}{12}, \bibinfo{number}{4} (\bibinfo{year}{2022}), \bibinfo{pages}{1--26}.
\newblock


\bibitem[Feng et~al\mbox{.}(2023b)]%
        {feng2023towards}
\bibfield{author}{\bibinfo{person}{Sidong Feng}, \bibinfo{person}{Haochuan Lu}, \bibinfo{person}{Ting Xiong}, \bibinfo{person}{Yuetang Deng}, {and} \bibinfo{person}{Chunyang Chen}.} \bibinfo{year}{2023}\natexlab{b}.
\newblock \showarticletitle{Towards Efficient Record and Replay: A Case Study in WeChat}. In \bibinfo{booktitle}{\emph{Proceedings of the 31st ACM Joint European Software Engineering Conference and Symposium on the Foundations of Software Engineering}}. \bibinfo{pages}{1681--1692}.
\newblock


\bibitem[Feng et~al\mbox{.}(2024)]%
        {feng2024mud}
\bibfield{author}{\bibinfo{person}{Sidong Feng}, \bibinfo{person}{Suyu Ma}, \bibinfo{person}{Han Wang}, \bibinfo{person}{David Kong}, {and} \bibinfo{person}{Chunyang Chen}.} \bibinfo{year}{2024}\natexlab{}.
\newblock \showarticletitle{MUD: Towards a Large-Scale and Noise-Filtered UI Dataset for Modern Style UI Modeling}. In \bibinfo{booktitle}{\emph{Proceedings of the CHI Conference on Human Factors in Computing Systems}}. \bibinfo{pages}{1--14}.
\newblock


\bibitem[Feng et~al\mbox{.}(2021)]%
        {feng2021auto}
\bibfield{author}{\bibinfo{person}{Sidong Feng}, \bibinfo{person}{Suyu Ma}, \bibinfo{person}{Jinzhong Yu}, \bibinfo{person}{Chunyang Chen}, \bibinfo{person}{Tingting Zhou}, {and} \bibinfo{person}{Yankun Zhen}.} \bibinfo{year}{2021}\natexlab{}.
\newblock \showarticletitle{Auto-icon: An automated code generation tool for icon designs assisting in ui development}. In \bibinfo{booktitle}{\emph{Proceedings of the 26th International Conference on Intelligent User Interfaces}}. \bibinfo{pages}{59--69}.
\newblock


\bibitem[Feng et~al\mbox{.}(2023c)]%
        {feng2023efficiency}
\bibfield{author}{\bibinfo{person}{Sidong Feng}, \bibinfo{person}{Mulong Xie}, {and} \bibinfo{person}{Chunyang Chen}.} \bibinfo{year}{2023}\natexlab{c}.
\newblock \showarticletitle{Efficiency matters: Speeding up automated testing with gui rendering inference}. In \bibinfo{booktitle}{\emph{2023 IEEE/ACM 45th International Conference on Software Engineering (ICSE)}}. IEEE, \bibinfo{pages}{906--918}.
\newblock


\bibitem[Feng et~al\mbox{.}(2023d)]%
        {feng2023read}
\bibfield{author}{\bibinfo{person}{Sidong Feng}, \bibinfo{person}{Mulong Xie}, \bibinfo{person}{Yinxing Xue}, {and} \bibinfo{person}{Chunyang Chen}.} \bibinfo{year}{2023}\natexlab{d}.
\newblock \showarticletitle{Read It, Don't Watch It: Captioning Bug Recordings Automatically}. In \bibinfo{booktitle}{\emph{2023 IEEE/ACM 45th International Conference on Software Engineering (ICSE)}}. IEEE, \bibinfo{pages}{2349--2361}.
\newblock


\bibitem[Gao et~al\mbox{.}(2019)]%
        {gao2019emerging}
\bibfield{author}{\bibinfo{person}{Cuiyun Gao}, \bibinfo{person}{Wujie Zheng}, \bibinfo{person}{Yuetang Deng}, \bibinfo{person}{David Lo}, \bibinfo{person}{Jichuan Zeng}, \bibinfo{person}{Michael~R Lyu}, {and} \bibinfo{person}{Irwin King}.} \bibinfo{year}{2019}\natexlab{}.
\newblock \showarticletitle{Emerging app issue identification from user feedback: Experience on wechat}. In \bibinfo{booktitle}{\emph{2019 IEEE/ACM 41st International Conference on Software Engineering: Software Engineering in Practice (ICSE-SEIP)}}. IEEE, \bibinfo{pages}{279--288}.
\newblock


\bibitem[Gomez et~al\mbox{.}(2013)]%
        {gomez2013reran}
\bibfield{author}{\bibinfo{person}{Lorenzo Gomez}, \bibinfo{person}{Iulian Neamtiu}, \bibinfo{person}{Tanzirul Azim}, {and} \bibinfo{person}{Todd Millstein}.} \bibinfo{year}{2013}\natexlab{}.
\newblock \showarticletitle{Reran: Timing-and touch-sensitive record and replay for android}. In \bibinfo{booktitle}{\emph{2013 35th International Conference on Software Engineering (ICSE)}}. IEEE, \bibinfo{pages}{72--81}.
\newblock


\bibitem[Gu et~al\mbox{.}(2019)]%
        {gu2019practical}
\bibfield{author}{\bibinfo{person}{Tianxiao Gu}, \bibinfo{person}{Chengnian Sun}, \bibinfo{person}{Xiaoxing Ma}, \bibinfo{person}{Chun Cao}, \bibinfo{person}{Chang Xu}, \bibinfo{person}{Yuan Yao}, \bibinfo{person}{Qirun Zhang}, \bibinfo{person}{Jian Lu}, {and} \bibinfo{person}{Zhendong Su}.} \bibinfo{year}{2019}\natexlab{}.
\newblock \showarticletitle{Practical GUI testing of Android applications via model abstraction and refinement}. In \bibinfo{booktitle}{\emph{2019 IEEE/ACM 41st International Conference on Software Engineering (ICSE)}}. IEEE, \bibinfo{pages}{269--280}.
\newblock


\bibitem[Guo et~al\mbox{.}(2019)]%
        {guo2019sara}
\bibfield{author}{\bibinfo{person}{Jiaqi Guo}, \bibinfo{person}{Shuyue Li}, \bibinfo{person}{Jian-Guang Lou}, \bibinfo{person}{Zijiang Yang}, {and} \bibinfo{person}{Ting Liu}.} \bibinfo{year}{2019}\natexlab{}.
\newblock \showarticletitle{Sara: self-replay augmented record and replay for Android in industrial cases}. In \bibinfo{booktitle}{\emph{Proceedings of the 28th acm sigsoft international symposium on software testing and analysis}}. \bibinfo{pages}{90--100}.
\newblock


\bibitem[Hu et~al\mbox{.}(2018)]%
        {hu2018appflow}
\bibfield{author}{\bibinfo{person}{Gang Hu}, \bibinfo{person}{Linjie Zhu}, {and} \bibinfo{person}{Junfeng Yang}.} \bibinfo{year}{2018}\natexlab{}.
\newblock \showarticletitle{AppFlow: using machine learning to synthesize robust, reusable UI tests}. In \bibinfo{booktitle}{\emph{Proceedings of the 2018 26th ACM Joint Meeting on European Software Engineering Conference and Symposium on the Foundations of Software Engineering}}. \bibinfo{pages}{269--282}.
\newblock


\bibitem[Kaufman et~al\mbox{.}(2012)]%
        {kaufman2012leakage}
\bibfield{author}{\bibinfo{person}{Shachar Kaufman}, \bibinfo{person}{Saharon Rosset}, \bibinfo{person}{Claudia Perlich}, {and} \bibinfo{person}{Ori Stitelman}.} \bibinfo{year}{2012}\natexlab{}.
\newblock \showarticletitle{Leakage in data mining: Formulation, detection, and avoidance}.
\newblock \bibinfo{journal}{\emph{ACM Transactions on Knowledge Discovery from Data (TKDD)}} \bibinfo{volume}{6}, \bibinfo{number}{4} (\bibinfo{year}{2012}), \bibinfo{pages}{1--21}.
\newblock


\bibitem[Li et~al\mbox{.}(2020)]%
        {li2020mapping}
\bibfield{author}{\bibinfo{person}{Yang Li}, \bibinfo{person}{Jiacong He}, \bibinfo{person}{Xin Zhou}, \bibinfo{person}{Yuan Zhang}, {and} \bibinfo{person}{Jason Baldridge}.} \bibinfo{year}{2020}\natexlab{}.
\newblock \showarticletitle{Mapping natural language instructions to mobile UI action sequences}.
\newblock \bibinfo{journal}{\emph{arXiv preprint arXiv:2005.03776}} (\bibinfo{year}{2020}).
\newblock


\bibitem[Mao et~al\mbox{.}(2016)]%
        {mao2016sapienz}
\bibfield{author}{\bibinfo{person}{Ke Mao}, \bibinfo{person}{Mark Harman}, {and} \bibinfo{person}{Yue Jia}.} \bibinfo{year}{2016}\natexlab{}.
\newblock \showarticletitle{Sapienz: Multi-objective automated testing for android applications}. In \bibinfo{booktitle}{\emph{Proceedings of the 25th International Symposium on Software Testing and Analysis}}. \bibinfo{pages}{94--105}.
\newblock


\bibitem[Raffel et~al\mbox{.}(2020)]%
        {raffel2020exploring}
\bibfield{author}{\bibinfo{person}{Colin Raffel}, \bibinfo{person}{Noam Shazeer}, \bibinfo{person}{Adam Roberts}, \bibinfo{person}{Katherine Lee}, \bibinfo{person}{Sharan Narang}, \bibinfo{person}{Michael Matena}, \bibinfo{person}{Yanqi Zhou}, \bibinfo{person}{Wei Li}, {and} \bibinfo{person}{Peter~J Liu}.} \bibinfo{year}{2020}\natexlab{}.
\newblock \showarticletitle{Exploring the limits of transfer learning with a unified text-to-text transformer}.
\newblock \bibinfo{journal}{\emph{The Journal of Machine Learning Research}} \bibinfo{volume}{21}, \bibinfo{number}{1} (\bibinfo{year}{2020}), \bibinfo{pages}{5485--5551}.
\newblock


\bibitem[Singhal et~al\mbox{.}(2001)]%
        {singhal2001modern}
\bibfield{author}{\bibinfo{person}{Amit Singhal} {et~al\mbox{.}}} \bibinfo{year}{2001}\natexlab{}.
\newblock \showarticletitle{Modern information retrieval: A brief overview}.
\newblock \bibinfo{journal}{\emph{IEEE Data Eng. Bull.}} \bibinfo{volume}{24}, \bibinfo{number}{4} (\bibinfo{year}{2001}), \bibinfo{pages}{35--43}.
\newblock


\bibitem[Su et~al\mbox{.}(2017)]%
        {su2017guided}
\bibfield{author}{\bibinfo{person}{Ting Su}, \bibinfo{person}{Guozhu Meng}, \bibinfo{person}{Yuting Chen}, \bibinfo{person}{Ke Wu}, \bibinfo{person}{Weiming Yang}, \bibinfo{person}{Yao Yao}, \bibinfo{person}{Geguang Pu}, \bibinfo{person}{Yang Liu}, {and} \bibinfo{person}{Zhendong Su}.} \bibinfo{year}{2017}\natexlab{}.
\newblock \showarticletitle{Guided, stochastic model-based GUI testing of Android apps}. In \bibinfo{booktitle}{\emph{Proceedings of the 2017 11th Joint Meeting on Foundations of Software Engineering}}. \bibinfo{pages}{245--256}.
\newblock


\bibitem[Touvron et~al\mbox{.}(2023)]%
        {touvron2023llama}
\bibfield{author}{\bibinfo{person}{Hugo Touvron}, \bibinfo{person}{Thibaut Lavril}, \bibinfo{person}{Gautier Izacard}, \bibinfo{person}{Xavier Martinet}, \bibinfo{person}{Marie-Anne Lachaux}, \bibinfo{person}{Timoth{\'e}e Lacroix}, \bibinfo{person}{Baptiste Rozi{\`e}re}, \bibinfo{person}{Naman Goyal}, \bibinfo{person}{Eric Hambro}, \bibinfo{person}{Faisal Azhar}, {et~al\mbox{.}}} \bibinfo{year}{2023}\natexlab{}.
\newblock \showarticletitle{Llama: Open and efficient foundation language models}.
\newblock \bibinfo{journal}{\emph{arXiv preprint arXiv:2302.13971}} (\bibinfo{year}{2023}).
\newblock


\bibitem[Wang et~al\mbox{.}(2024)]%
        {wang2024feedback}
\bibfield{author}{\bibinfo{person}{Dingbang Wang}, \bibinfo{person}{Yu Zhao}, \bibinfo{person}{Sidong Feng}, \bibinfo{person}{Zhaoxu Zhang}, \bibinfo{person}{William~GJ Halfond}, \bibinfo{person}{Chunyang Chen}, \bibinfo{person}{Xiaoxia Sun}, \bibinfo{person}{Jiangfan Shi}, {and} \bibinfo{person}{Tingting Yu}.} \bibinfo{year}{2024}\natexlab{}.
\newblock \showarticletitle{Feedback-Driven Automated Whole Bug Report Reproduction for Android Apps}.
\newblock \bibinfo{journal}{\emph{arXiv preprint arXiv:2407.05165}} (\bibinfo{year}{2024}).
\newblock


\bibitem[Xie et~al\mbox{.}(2020)]%
        {xie2020uied}
\bibfield{author}{\bibinfo{person}{Mulong Xie}, \bibinfo{person}{Sidong Feng}, \bibinfo{person}{Zhenchang Xing}, \bibinfo{person}{Jieshan Chen}, {and} \bibinfo{person}{Chunyang Chen}.} \bibinfo{year}{2020}\natexlab{}.
\newblock \showarticletitle{UIED: a hybrid tool for GUI element detection}. In \bibinfo{booktitle}{\emph{Proceedings of the 28th ACM Joint Meeting on European Software Engineering Conference and Symposium on the Foundations of Software Engineering}}. \bibinfo{pages}{1655--1659}.
\newblock


\bibitem[Xie et~al\mbox{.}(2022)]%
        {xie2022psychologically}
\bibfield{author}{\bibinfo{person}{Mulong Xie}, \bibinfo{person}{Zhenchang Xing}, \bibinfo{person}{Sidong Feng}, \bibinfo{person}{Xiwei Xu}, \bibinfo{person}{Liming Zhu}, {and} \bibinfo{person}{Chunyang Chen}.} \bibinfo{year}{2022}\natexlab{}.
\newblock \showarticletitle{Psychologically-inspired, unsupervised inference of perceptual groups of GUI widgets from GUI images}. In \bibinfo{booktitle}{\emph{Proceedings of the 30th ACM Joint European Software Engineering Conference and Symposium on the Foundations of Software Engineering}}. \bibinfo{pages}{332--343}.
\newblock


\bibitem[Zadgaonkar(2013)]%
        {zadgaonkar2013robotium}
\bibfield{author}{\bibinfo{person}{Hrushikesh Zadgaonkar}.} \bibinfo{year}{2013}\natexlab{}.
\newblock \bibinfo{booktitle}{\emph{Robotium automated testing for android}}.
\newblock \bibinfo{publisher}{Packt Publishing}.
\newblock


\end{thebibliography}

\end{document}